\documentstyle[12pt,epsf]{article}
\newlength{\abstractwidth}
\renewcommand{\thefootnote}{\fnsymbol{footnote}}
\renewcommand{\thanks}[1]{\footnote{#1}} 
\newcommand{\starttext}{
\setcounter{footnote}{0}
\renewcommand{\thefootnote}{\arabic{footnote}}}
\renewcommand{\theequation}{\thesection.\arabic{equation}}
\newcommand{\be}{\begin{equation}}
\newcommand{\bea}{\begin{eqnarray}}
\newcommand{\eea}{\end{eqnarray}}
\newcommand{\beq}{\begin{equation}}
\newcommand{\ee}{\end{equation}}
\newcommand{\eeq}{\end{equation}}
\renewcommand{\a}{\alpha}
\renewcommand{\b}{\beta}
\newcommand{\g}{\gamma}

\def\nn{\nonumber}
\def\ket#1{\left|#1\right>}
\def\bra#1{\left<#1\right|}
\def\noket#1{\left.#1\right>}

\begin{document}
\renewcommand{\theequation}{\thesection.\arabic{equation}}
\begin{titlepage}
\bigskip
\rightline{} \rightline{SU-ITP 01-03} \rightline{hep-th/0103012}

\bigskip\bigskip\bigskip\bigskip

\begin{center}
{\Large \bf {Fundamentals on the Noncommutative Plane}}

\vskip 1 cm
{\it Yonatan Zunger$^\dagger$}
\vskip.3cm
Department of Physics

Stanford University

Stanford, CA 94305-4060

\end{center}

\vskip 1cm

\begin{abstract}

We consider the addition of charged matter (``fundametals'') to noncommutative
Yang-Mills theory and noncommutative QED, derive Feynman rules and tree-level
potentials for them, and study the divergence structure of the theory. These
particles behave very much as they do in the commutative theory, except that
(1) they occupy bound-state wavefunctions which are essentially those of
charged particles in magnetic fields, and (2) there is slight momentum
nonconservation at vertices.  There is no reduction in the degree of divergence
of charged fermion loops like that which affects nonplanar noncommutative
Yang-Mills diagrams.
\end{abstract}

\vfill
\footnoterule
\noindent
{\footnotesize
$\phantom{1}^\dagger$zunger@itp.stanford.edu}
\end{titlepage}

\starttext \baselineskip=18pt \setcounter{footnote}{0}

\setcounter{equation}{0}

The dynamics of dipoles in two dimensions in a strong background magnetic
field is governed by noncommutative Yang-Mills, (NCYM) \cite{SeiWitt, LennyNew,
BigSus, Sheikh} where the dipoles are described by the gauge fields which have
a noncommutative multiplication rule. This rule is equivalent to replacing
the coordinate functions $x$ and $y$ with operators that have certain commutation
relations, and writing the gauge fields as Taylor series in these operators. \cite{NonSol}
Thus the gauge fields are operator-valued quantities.

Given a set of operator-valued fields, it is natural to ask what the fields
are which would correspond to column vectors. In this paper we will introduce
such fields, called ``fundamentals'' since they carry the fundamental representation
of the operator algebra, and study their dynamics. We will work out Feynman rules, study
some basic processes such as their 1-loop mass renormalization and their Coulomb potential,
and examine the divergence structure of their theory.

Of course, such fields would not be worth studying if they did not occur in a physics
context. Fortunately they are easy to come across. We will begin by setting up the
problem of a charged particle in a magnetic field in the traditional quantum way, and
show that the gauge field emerges as an operator-valued quantity (in the usual way) and
that these fundamental fields naturally emerge as the description of the individual
charged particles. It is therefore natural to suspect that in more complicated systems
governed by noncommutative Yang-Mills theory (NCYM) these fields will show up as the
basic kind of charged matter. More recently such fields emerged in the study of
noncommutative Chern-Simons theory in 2-dimensional systems with boundary; \cite{Polychr}
there these fields were the boundary excitations of the theory. Such theories are
potentially very interesting as simple examples of holographic systems, and so their
boundary fields merit particular interest.

\section{Introduction: Noncommutative Geometry and Fundamentals}

Consider a charged nonrelativistic particle moving on a plane in an
applied constant magnetic field perpendicular to the plane. This system
satisfies the Schr\"odinger equation $H\ket{\psi}=i\hbar\partial_0\ket{\psi}$
with Hamiltonian
\be
H = \frac{1}{2m}\left(p-gA\right)^2
\label{eq:basehamilt}
\ee
where $A=\frac{1}{2}B(-y, x)$. Later on we will allow $A$ to fluctuate
about this value. This Hamiltonian really has two kinds of
fields in it: Hilbert-space-valued (``fundamental'') fields such as the
wave function, and operator-valued fields such as the vector potential.
In a moment we will make a change of variables which will simplify this
Hamiltonian at the expense of replacing the coordinates $x$ with
noncommuting coordinates $\tilde{x}$. The operator-valued fields such as
$A$ become functions of these noncommuting coordinates, and their dynamics
is fairly well-understood. Our objective here is to study how
the Hilbert space-valued fields behave both in this system and in the
broader context of noncommutative Yang-Mills theory. Following the
terminology from the mathematical literature, we refer to these as
fundamental fields. This name comes from the fact that if one writes
operators as matrices they carry two indices, while these fields carry
a single index; more technically, these fields carry the fundamental
representation of the algebra of functions on the space.

The change of variables is as follows. The quantity in the parentheses
of (\ref{eq:basehamilt}) is defined to be the canonical momentum $\tilde{p}$.
It obeys
\be
\left[\tilde{p}_x, \tilde{p}_y\right] = \left[p_x+\frac{1}{2}gBy,
p_y-\frac{1}{2}gBx\right] = igB\ .
\ee
We can define coordinates which are conjugate to these momenta in the
sense that $[\tilde{x}^i, \tilde{p}_j]=\delta^i_j$. This is satisfied by
\bea
\tilde{x} &=& -\frac{1}{gB}\tilde{p}_y \nn \\
\tilde{y} &=& \frac{1}{gB}\tilde{p}_x\ .
\eea
However, these coordinates do not commute with one another; they satisfy
\be
[\tilde{x}^i, \tilde{x}^j] = \epsilon^{ij} \frac{i}{gB} \equiv i\theta^{ij}\ .
\label{eq:xcom}
\ee

One should be aware at this point that the commutative limit of this theory
is not $\theta\rightarrow 0$ but $\theta\rightarrow\infty$. This is clear
if one substitutes into (\ref{eq:basehamilt}), since then the $\tilde p$'s
turn back into commuting $p$'s. The Schr\"odinger equation for this
Hamiltonian is
\be
H\ket{\psi} = \frac{1}{2m}\left(\tilde{p}_x^2+\tilde{p}_y^2\right)\ket{\psi} =
\frac{1}{2m}\left(\tilde{p}_x^2+\theta^{-2}\tilde{x}^2\right)\ket{\psi} =
E\ket{\psi}
\ee
and its solutions are 1D simple harmonic oscillator wave functions; but in
the limit $\theta\rightarrow\infty$ the eigenfunctions become degenerate and
can be brought by a change of basis back to 2D planewaves, which are the
solutions of the original equation (\ref{eq:basehamilt}) at infinite $\theta$.

The potentially confusing equation in the commutative limit is (\ref{eq:xcom}).
This equation is not disastrous simply because $\tilde{x}$ and $\tilde{y}$
do not have smooth limits as $\theta\rightarrow\infty$; they both diverge
linearly. In this limit those are simply no longer good coordinates. One
could, of course, make a change to a different set of coordinates where
$\tilde{x}$ and $\tilde{y}$ remain healthy in this limit (and their commutator
goes to zero) but in this case their conjugate momenta will
diverge.\footnote{The intuition for this is that in the noncommutative case,
both functions and derivatives live in the same algebra, but in the
commutative limit derivatives are not functions. So in this limit one
expects that one or the other will in some way fail to converge to a
good value. Rescaling so that the $\tilde{x}$'s remain good into the
commutative limit is somewhat more intuitive from the perspective of looking
at large $\theta$, but would make some of the field theory expressions
below appear unusual.}

\medskip

Given this set of ``coordinate'' operators, we can verify that they indeed
behave somewhat like ordinary coordinates on a space by defining derivatives
with respect to them. We use the canonical formula
\be
D_i = i \tilde{p}_i = i\theta^{-1}_{ij}\tilde{x}^j
\ee
$D_i$ generates derivatives on states in the Hilbert space by multiplication
and on operators by commutation. The first statement means that for
$\ket{x}$ an eigenket of $\tilde{x}$ and $dx$ infinitesimal,
\be
\tilde{x} \left(1+dx D_x\right) \ket{x} = x+dx(1+D_x \tilde{x})\ket{x} =
(x+dx)(1+dx D_x)\ket{x}\ ,
\ee
which implies that $(1+dx D_x)\ket{x}=\ket{x+dx}$, so
\be
D\ket{x} = \lim_{\Delta x \rightarrow 0} \frac{\ket{x+\Delta x} - \ket{x}}
{\Delta x}\ .
\ee
(This is just the usual proof that momenta generate translations) The
second statement follows from
\be
[D_i, \tilde{x}^j] = \delta_i^j\ ,
\ee
which implies that
\be
\left[D_x, \sum a_{mn}\tilde{x}^m \tilde{y}^n\right] = \sum m a_{mn}
\tilde{x}^{m-1} \tilde{y}^n\ ,
\label{eq:derivrule}
\ee
and the property
\be
[D, AB] = [D,A]B + A[D,B]
\ee
which holds for any operators $A$ and $B$. Together these mean that
$[D,\cdot]$ takes the derivative of any locally polynomial function of the
$\tilde{x}^i$ in the normal manner.

It may appear strange at first glance to say that $D$ acts on operators
by conjugation, since one typically writes operators as acting on one another
by matrix multiplication. What this statement really means is that for any
operator $A$ and any ket $\ket{\psi}$,
\be
DA\ket{\psi} = [D,A]\ket{\psi} + A D\ket{\psi}
\label{eq:compatcond}
\ee
which is the product rule for an operator and a ket.

\medskip

The combination of an algebra of operators (the set of operator-valued
functions and their multiplication law), a Hilbert space and a derivative
operator is the definition in algebraic geometry of a noncommutative space.
In the commutative limit, the algebra of operators turns into the algebra
of functions on the space, while the Hilbert space turns into the vector
space of $L^2$ functions.

This notion of noncommutative space is very closely related to the one that is
commonly seen in the physics literature. Typically noncommutative
field theories are defined by taking ordinary field theories and replacing
the product of two fields with the star product
\be
f(x)\star g(x) = \left.e^{-\frac{i}{2}\theta^{ij}\frac{\partial}{\partial y^i}
\frac{\partial}{\partial z^j}} f(y) g(z)\right|_{y=z=x}\ .
\label{eq:moyal}
\ee
This star product is exactly the multiplication rule one obtains from taking
the Taylor expansions of $f$ and $g$ and replacing the coordinates $x$ and
$y$ with the noncommuting coordinates $\tilde{x}$ and $\tilde{y}$ above:
\be
f(x) = \sum a_{mn} x^m y^n \longrightarrow \sum a_{mn}W(\tilde{x}^m, \tilde{y}^n)
\ee
where $W(\tilde{x}^m,\tilde{y}^n)$ represents the Weyl-ordered product of
the operators.

We have gone through this somewhat more circuitous operator-based derivation of
noncommutative theories since it highlights the presence of fundamental
fields and simultaneously makes it clear what needs to be done in order
to integrate them into noncommutative field theories. We wish to look at
these fields as models of particles charged under noncommutative QED or YM.

Before plunging into this subject, it is useful to look briefly at theories
of operator-valued fields. The theory of greatest interest to us is
noncommutative QED, which follows from taking the ordinary QED Lagrangian in
our background:
\be
{\cal L} = \frac{1}{4} F_{\mu\nu} F^{\mu\nu}
\label{eq:qed}
\ee
where $\mu,\nu=0,1,2$ and
\be
F_{\mu\nu} = \frac{1}{ig}\left[\partial_\mu -ig A^0_\mu -ig A_\mu,
\partial_\nu -ig A^0_\nu -ig A_\nu\right]\ .
\ee
$A^0_\mu$ is the background vector potential $A$ of (\ref{eq:basehamilt}),
and the $\partial_\mu$ are ordinary partial derivatives. But writing these
derivatives in terms of momenta, we can clearly combine them with the
$A^0$ to form our $\tilde{p}_\mu$. (Since $A^0_0=0$, $\tilde{p}_0=p_0$
and the time components are unchanged) Therefore we can rewrite the field
strength as
\be
F_{\mu\nu} = \frac{1}{ig}\left[D_\mu -ig A_\mu, D_\nu-ig A_\nu\right]\ .
\ee
As we noted above, this is exactly the field strength one would get if one
replaced multiplication with the star product in (\ref{eq:qed}). The resulting
field theory is noncommutative Yang-Mills.

One point about the solution to this theory, which will be useful later,
is that the eigenfunctions of a free boson (the solutions to the Klein-Gordon
equation) are essentially the same in noncommutative space and commutative
space. The equation is
\be
\partial_0^2 A_\mu - \delta^{ij}[D_i, [D_j, A_\mu]] = 0
\ee
and by writing $A_\mu$ as a Taylor series in the $\tilde{x}$ one easily
verifies that the solutions are
\be
A_\mu = e^{i(\omega t - q\cdot \hat{x})}
\ee
where $\omega^2-q^2=0$.\footnote{The fact that the solutions to the
Klein-Gordon equation have the same form as in commutative space is not
generic; it depends crucially on the fact that the commutator of two
momenta is proportional to the identity. For more general noncommutative
spaces the solution would be more complicated.}

We now turn to our main subject of interest, the fundamental fields which
are minimally coupled to the gauge fields of NCYM.
Because both of the cases of physical interest to us (the particle in the
magnetic field and string endings on a brane) have spin $1/2$, we will spend
most of our time examining solutions of the Dirac equation
\be
\left(i\not\partial + m\right)\ket{\psi} = 0\ .
\label{eq:diraceqn}
\ee
There are some technical subtleties with the case of finite mass, since in a
non-Lorentz-invariant theory (Lorentz-invariance is broken by the magnetic
field) these fermions will lead to the violation of Ward identities and
possible sicknesses in the theory at higher loops. As such we will begin our
analysis by considering the limit where there are no kinetic terms in the
Lagrangian. ($m=\infty$, the extreme nonrelativistic limit) This will allow
us to study some simple effects like electron mass renormalization and the
Coulomb force. At the end of this paper we will discuss the divergence structure
of the theory, and at that point it will be at least formally interesting to
restore the space derivatives in (\ref{eq:diraceqn}). We will do this,
although we reserve judgement on the physical relevance of such solutions.

\section{The Dirac action in the nonrelativistic limit}

For the remainder of this paper, we will look exclusively at the properties
of systems in noncommutative space, so where there is no risk of confusion
we will drop tildes on the $x$'s and the $p$'s and replace $D$ with $\partial$.

We would like to write the Dirac action for Hilbert space-valued fermions in
noncommutative space. If space satisfies
\be
[x^i, x^j]=i\theta^{ij}=i\theta \epsilon^{ij}
\label{eq:comrel}
\ee
where $i,j$ are space indices and $\theta$ is a real number, then space
derivatives are given by
\be
\partial_i = i\theta^{-1}_{ij}x^j\ .
\ee
Time derivatives are unchanged from the commutative case. This means that
the Laplacian operator is
\be
\nabla^2 \sim \partial_0^2 + \theta^{-2} x^2
\ee
which is not Lorentz-invariant.

For the reasons discussed above, we will consider states that have no
space derivatives on them, and so write an action
\be
S = \int dt d^2 x \bar\psi\left(i\gamma^0 \nabla_0 + m\right)\psi + L_{YM}\ .
\ee
$\nabla$ here is a gauge-covariant derivative coupling to an electromagnetic
field:
\be
\nabla = \partial - igA\ .
\ee
This action has a gauge symmetry $\psi\rightarrow (1+i\alpha)\psi$,
$A^\mu\rightarrow A^\mu - \nabla^\mu\alpha$, where $\alpha$ is an
infinitesimal operator-valued quantity.

We first get the fermion propagator and eigenstates by solving the system
when $g=0$. (No coupling constant) Then the Green's function satisfies
\be
\left(i\gamma^0\partial_0+m\right)G=i\delta(x) \Rightarrow G(k)=\frac{i}
{\gamma^0 k_0+m}\ .
\ee
This is our fermion propagator. The fermion eigenstates follow from solving
the Dirac equation
\be
\left(i\gamma^0 \partial_0+m\right)\psi=0 \Rightarrow \psi(t,z)=
e^{\pm imt}\left(\begin{array}{c}\ket{\psi_\uparrow}\\\ket{\psi_\downarrow}
\end{array}\right)\ .
\label{eq:nospdirac}
\ee
Since there is no space dependence of the Dirac equation, the
$\ket{\psi_{\uparrow,\downarrow}}$ are unconstrained and so we are free to
choose any basis of square-normalizable functions for them.
We will label these by $\ket{n\alpha\pm}$, where $n$ is an index, $\alpha$
is the spin state and $\pm$ is the sign of the frequency in
(\ref{eq:nospdirac}).  (We will choose a specific basis later when we consider
the full Dirac equation; for now any basis for the Hilbert space will do)
The completeness relationship for these kets is
\be
\sum_n \ket{n\a\pm}\bra{n\a'\pm'} = \delta_{\a\a'}\delta_{\pm\pm'}
\ee
since the $\ket{\psi_\uparrow}$ and $\ket{\psi_\downarrow}$ are independently
arbitrary members of the Hilbert space. More generally, we can define
off-shell states $\ket{z_0;n\a\pm}$ which are also labelled by a frequency
$z_0$ which may differ from $m$. These show up in sums over intermediate
states. Their completeness relationship is
\be
\int \frac{dz_0}{2\pi} \sum_n \ket{z_0;n\a\pm}\bra{z_0;n\a'\pm'} =
\delta_{\a\a'}\delta_{\pm\pm'}\ .
\ee

\begin{figure}[ht]
\centerline{\epsffile{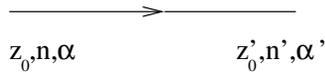}}
\caption{The electron propagator}
\label{fig:eprop}
\end{figure}

Using this basis, the electron propagator between two off-shell states
is (figure~\ref{fig:eprop})
\be
\frac{i}{\gamma^0 z_0+m}\bra{n'\a'\pm}\noket{n\a\pm} (2\pi) \delta(z'_0-z_0)
= \frac{i}{\gamma^0 z_0 +m} (2\pi) \delta(z'_0-z_0)\delta_{nn'}\delta_{\a\a'}\ .
\ee

External fermion legs are associated with factors of the on-shell fermion wave
function $\ket{n\alpha\pm}$ times $\sqrt{2m}$. (This factor is for consistency
with standard normalizations in field theory)

\begin{figure}[ht]
\centerline{\epsffile{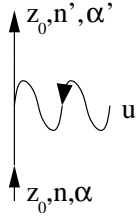}}
\caption{The electron-photon vertex}
\label{fig:vertex}
\end{figure}

Finally, let $g\not=0$ and let us find the electron-photon vertex. This
follows most easily from taking the variation of the action using the usual
LSZ formulation, but it is instructive to get it from time-dependent
perturbation theory as well. We apply a time-dependent field
\be
A^0 = \eta(t) e^{-i(u_1 x_1 + u_2 x_2)} \equiv \eta(t) G(u)
\ee
where $x_{1,2}$ are the coordinates of the noncommutative plane and $u_{1,2}$
are $c$-numbers.
The perturbation in the Hamiltonian is simply $\Delta H = gA^0$ (from
minimal coupling in the nonrelativistic theory; that's ok because since
there are no space derivatives we are essentially always working in the
nonrelativistic limit) Then from ordinary perturbation theory we get
\be
\psi(n, t) = \sum_m \frac{1}{m!}\left(\frac{-ig}{\hbar}\int dt \eta(t)\right)^m
\sum_{n_1\cdots n_{m-1}} \bra{n}G(u)\ket{n_1}\bra{n_1}G(u)\ket{n_2}\cdots
\bra{n_{m-1}}G(u)\ket{n_0}
\ee
if the particle is initially in the state $\left|n_0\right>$. This corresponds
to an insertion of $-ig$ at each vertex, which is the same as the result which
one finds from varying the Lagrangian.
However, the momentum-conserving delta function which usually comes from the
inner product is changed due to factors of $G(u)$. The combined vertex
(including momentum factors) is thus
\be
-ig (2\pi)\delta(z'_0-z_0-u_0) \bra{z'_0;n'\a'\pm}G(u)\ket{z_0;n\a\pm} \ .
\label{eq:pspsgvert}
\ee

The matrix element in (\ref{eq:pspsgvert}) can most interestingly be evaluated
if we expand the fermion states in the (nonorthonormal) momentum
basis
\be
\left|z\right> = e^{-iz\cdot x}\ket{0} = e^{-i\sqrt{\frac{\theta}{2}}
\left(za+z^\star a^\dagger\right)}\ket{0} = G(z)\ket{0}
\label{eq:zdef}
\ee
where the $a$ and $a^\dagger$ are SHO ladder operators normalized such that
$[a,a^\dagger]=1$:
\be
a = \frac{1}{\sqrt{2\theta}}(x^1-ix^2)\ ,
\ee
so that
\be
\ket{n\alpha\pm} = \int d^2 z u_{n\alpha\pm}(z) \ket{z}\ .
\ee
Then the fermion-photon vertex between fermions in momentum eigenstates
$z$ and $z'$ is given by
\be
-ig (2\pi) \delta(z'_0-z_0-u_0) f_1(z,z';u) \delta_{\a\a'}
\ee
where
\be
f_1(z,z';u) = \left<z'\right|G(u)\left|z\right> = \left<0\right|G(-z')G(u)
G(z)\left|0\right>\ .
\label{eq:f1def}
\ee

This function can be evaluated using the Baker-Hausdorff lemma and
(\ref{eq:zdef}) to be
\be
f_1(z,z';u) = e^{-\frac{\theta}{4}\left(|q|^2 - \left(q^\star z'- q z'^\star
\right) - \left(u^\star z + u z^\star\right) \right)}
\label{eq:f1val}
\ee
where $q=z+u-z'$ is the momentum nonconservation (``wobble'') at the vertex.
Note that nonconservation is Gaussian damped and vanishes in the commutative
limit; as $\theta\rightarrow\infty$ this function approaches
$\delta^2(z'-z-u)$, justifying the interpretation of $z$ as a space momentum.
These phases are the analogue for fundamentals of the phase factors
$\exp[i\sum_{i<j}p_i\wedge p_j]$ which occur at gauge boson vertices in NCYM.
\cite{BigSus}

The momentum nonconservation follows from the fact that momenta along different
directions do not commute, and therefore the labelling of fermion states by
two components of momentum was mildly fallacious.

More generally, we can define
\be
f_n(z,z';p_1,\ldots p_n) = \bra{z'}G(p_1)\cdots G(p_n)\ket{z}
\ee
which can be easily evaluated using the Baker-Hausdorff lemma. That tells
us that
\be
\bra{0}G(p_1)\cdots G(p_N)\ket{0} = e^{-\frac{\theta}{4}|q|^2} e^{i
\sum_{j<i} p_i\wedge p_j}
\label{eq:fn0def}
\ee
where $q=\sum p_i$ and $p_1\wedge p_2 = \frac{1}{2}\theta_{\mu\nu} p_1^\mu
p_2^\nu$, (The factor of $2$ in the wedge product comes from our having
defined $[x^i,x^j]=i\theta^{ij}$, which differs by this factor of 2 from
that used in some other papers) and so
\be
f_n(z,z';p_1,\ldots p_n) = e^{-\frac{\theta}{4}|q+z-z'|^2}
e^{i\left(\sum_{j<i}p_i\wedge p_j + (z+z')\wedge q - z\wedge z'\right)}\ .
\label{eq:fndef}
\ee
This function will be useful in the calculation of general Feynman diagrams
later.

One should note, though, that we did not Fourier transform anywhere to cause
the space components of $z$ to behave as momenta, whereas we did do so for
the time component in solving the Dirac equation. This means that integrals
over momenta in loops have measure $d^3z/2\pi$, rather than $(2\pi)^3$. This
corresponds with the absence of a $(2\pi)^2$ in the $\theta\rightarrow\infty$
limit of $f_1$.

The remaining Feynman rules of the theory are simply those of noncommutative
Yang-Mills theory. One should note that only the zero component of the
gauge field couples to the fermions.

\section{The fermion two-point function}

\begin{figure}[ht]
\centerline{\epsffile{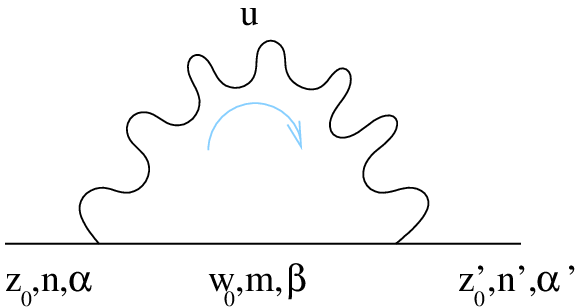}}
\label{fig:twopoint}
\end{figure}

The one-loop contribution to the fermion two-point function is given in
figure (\ref{fig:twopoint}).

Using the Feynman rules above, the correction to the propagator is given
by
\be
(-ig)^2\int \frac{d^3 u dw_0}{(2\pi)^4}\sum_{m\b} \bra{z'_0;n'\a'\pm}G(-u)
\ket{w_0;m\b\pm} \frac{i}{\gamma^0 w_0+m} \frac{i g_{00}}{u^2-\mu^2}
\bra{w_0;m\b\pm}G(u)\ket{z_0;n\a\pm}
\ee
where $\mu$ is a photon mass introduced as an infrared regulator. We can
perform the $m\beta$ sum immediately, since they do not appear in any
propagators;
\bea
\sum_{m\beta} \bra{z'_0;n'\a'\pm}G(-u)\ket{w_0;m\b\pm}\bra{w_0;m\b\pm}G(u)
\ket{z_0;n\a\pm} &=& \bra{z'_0;n'\a'\pm}G(-u)G(u)\ket{z_0;n\a\pm} \nn \\
&=& \bra{z'_0;n'\a'\pm}\noket{z_0;n\a\pm} \nn \\
&=& (2\pi)\delta(z'_0-z_0)\delta_{nn'}\delta_{\a\a'}
\eea
so the two-point function is
\be
(-ig)^2 \int \frac{d^3u}{(2\pi)^3} \frac{i}{\gamma^0(z_0-u_0) +m}
\frac{i g_{00}}{u^2-\mu^2} \delta(z'_0-z_0) \delta_{nn'}\delta_{\a\a'}
\ee
This is exactly the commutative result, except for the absence of space
momenta in the fermion denominator. (Corresponding to the nonrelativistic
limit we have taken)

Therefore due to a fortuitous cancellation the effects of noncommutativity --
and specifically, the smearing of momentum conservation -- cancels out of
this diagram. This is our first example of the insensitivity of fermion
properties to the value of $\theta$.

Incidentally, the integral above can be performed; using a hard cutoff
$\Lambda$ to regulate momenta in the UV gives
\be
\frac{g^2}{4\pi} \int_\mu^\Lambda du \frac{1}{z_0-u+m}\log \left.\frac
{u^2+\mu^2}{u^2-\Lambda^2+\mu^2}\right|_{\mu\rightarrow 0}\ .
\ee
This is the same as the two-point function in the commutative
nonrelativistic theory, and still has the characteristic log divergence.

\section{Scattering and the Two-Body Hamiltonian}

One simple process where we can see a nontrivial effect of momentum smearing is
in the scattering of two electrons off each other.

\begin{figure}[ht]
\centerline{\epsffile{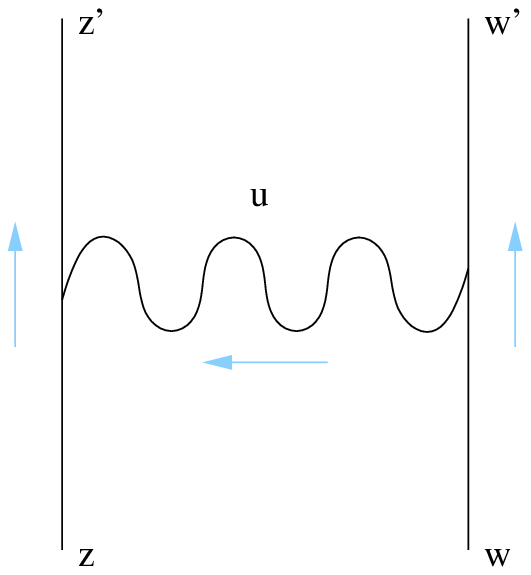}}
\label{eq:scatter}
\end{figure}

The amplitude is (n.b. we are using a mostly negative metric)
\bea
iM &=& (-ig)^2\int \frac{d^3 u}{2\pi} \bar{u}(z) f_1(z,z';u) u(z) \frac{i}{u^2}
\bar{u}(w) f_1(w',w;-u) u(w') \nonumber \\
&=& -2img^2\delta(z'_0+w'_0-z_0-w_0) \int d^2 u \frac{ \left<z\right|G(u)
\left|z'\right>\left<w\right|G(-u)\left|w'\right>}{u^2-(z'_0-z_0)^2}\ .
\eea
As $\theta\rightarrow\infty$, the expectation values in the numerator become
space-momentum-conserving $\delta$-functions, and this converges to the usual
result for scattering of particles in two dimensions. The different term
corresponds to the uncertainty in the momentum transfer.

We can extract the effective two-body Hamiltonian from this result using
the first Born approximation. Noting that the $f$'s arise from sandwiching
photon momentum operators $G(u)$ between external fermion states, the total
Hamiltonian is
\be
H = g^2 \int \frac{d^2 u}{|u|^2} G(u)\otimes G(-u)
\ee
which acts on the direct product Hilbert space of two fermions. Since these
are identical fermions, the final eigenfunctions must be antisymmetrized.

We can diagonalize this Hamiltonian in a straightforward manner. In terms
of SHO ladders for the two factors of the Hilbert space, it is equal to
\be
H = g^2\int \frac{d^2 u}{|u|^2}
e^{-i\sqrt{\frac{\theta}{2}}(u a + u^\star a^\dagger)}
e^{i\sqrt{\frac{\theta}{2}}(u b + u^\star b^\dagger)}
\ee
Since $a$ and $b$ commute it is useful to define a combined ladder $A=a-b$
and write
\be
H=g^2\int \frac{d^2 u}{|u|^2}
e^{-i\sqrt{\frac{\theta}{2}}(u A + u^\star A^\dagger)}
\label{eq:hreduced}
\ee

Now, two pairs of commuting SHO ladders naturally suggests the Schwinger
construction of angular momentum.\footnote{Thanks to L. Susskind for
suggesting the following construction.} We define the operators
\bea
N &=& a^\dagger a + b^\dagger b \nn \\
Q_z &=& \frac{1}{2}(a^\dagger b + b^\dagger a) \nn \\
Q_y &=& \frac{1}{2}(a^\dagger a - b^\dagger b) \nn \\
Q_x &=& \frac{i}{2}(b^\dagger a - a^\dagger b)\ .
\eea
The operators $Q_{x,y,z}$ generate the algebra $SU(2)$, and all commute with
the total number operator $N$. $N$ is related to the quadratic Casimir by
\be
Q^2 = \frac{1}{2}N\left(\frac{1}{2}N+1\right)\ .
\ee
The eigenstates of the paired SHO's (which are a basis for the Hilbert
space) may be decomposed into an angular-momentum basis
\be
\left|jm\right> = \frac{a^{\dagger(j+m)}b^{\dagger(j-m)}}{\sqrt{(j+m)!(j-m)!}}
\left|0 0\right>
\ee
where $\left|0 0\right>$ is the ground state of the SHO. These states are
eigenstates of $N$ and $Q_z$ with eigenvalues $2j$ and $m$, respectively, and
$j$ and $m$ take the usual values for angular-momentum eigenstates.

These operators are useful because the Hamiltonian commutes with both $N$ and
$Q_z$. The easiest way to see this is to expand the exponential in a Taylor
series and do the integral in polar coordinates. We introduce both upper
and lower bounds $\Lambda$ and $\mu$ on the radial component of this,
corresponding to UV and IR cutoffs respectively.

\bea
H &=& g^2 \int \frac{du d\alpha}{u} e^{-iu\sqrt{\frac{\theta}{2}}(e^{-i\alpha}
A^\dagger + e^{i\alpha} A)} \nn \\
&=& g^2 \sum_{N=0}^\infty \frac{1}{N!}\int d\alpha du u^{N-1}
\left(-i\sqrt{\frac{\theta}{2}}\right)^N
\left(e^{-i\alpha} A^\dagger + e^{i\alpha} A\right)^N \nn \\
&=& g^2 \int_0^{2\pi} d\alpha \left[\log \frac{\Lambda}{\mu} + \sum_{N=1}^\infty
\sum_{M=0}^N \frac{1}{N\cdot N!} \left(-i\Lambda\sqrt{\frac{\theta}{2}}
\right)^N e^{i(N-2M)\alpha} C(A^{\dagger N} A^{N-M}) \right]
\eea

$C(A^{\dagger N} A^{N-M})$ is defined to be the sum of all permutations of $N$
$A^\dagger$'s and $(N-M)$ $A$'s, treating permutations which differ by swapping
two identical elements as distinct. (So there are $N!$ terms) We have
implicitly let $\mu\rightarrow 0$ in the $N\ge 1$ terms since there are no
IR divergences there. The $\alpha$ integral then gives $2\pi\delta(2M-N)$,
so
\be
H = 2\pi g^2\left[\log\frac{\Lambda}{\mu} + \sum_{N=1}^\infty \frac{1}
{2N(2N)!} \left(-\frac{\Lambda^2\theta}{2}\right)^N C(A^{\dagger N} A^N)
\right]\ .
\ee

Now, the $C$ term can be reordered using the commutation relation $[A,A^\dagger]=2$ into a polynomial in $A^\dagger A$. The precise form of this polynomial is
messy and involves a good deal of combinatorics, but its leading term is
$(2N)!(A^\dagger A)^N$. (Lower terms are subdominant at large $N$ and so
become weak when $\theta\Lambda^2$ is large) So pulling out an overall
factor of $1/2$, we find
\be
H = \pi g^2\left[\log \frac{\Lambda^2}{\mu^2} + \sum \frac{1}{N}\left(
-\frac{\Lambda^2\theta}{2}A^\dagger A\right)^N + corrxns.\right]
\ee

At this point one should note that, although the integral is an infinite
series in $\Lambda$, it is only logarithmically divergent. This can be seen
from the original integral since it is given by a phase factor divided by
$u$. In fact, if one replaces $A$ and $A^\dagger$ in (\ref{eq:hreduced})
by numbers (as they would be in matrix elements of the Hamiltonian) the
integral can be done explicitly in terms of Bessel functions, and apart
from the overall logarithm term the integral goes to a constant in the
large-$\Lambda$ limit. The entire $\Lambda$-dependence is therefore limited
to replacing $\mu$ and $\theta$ with the dimensionless
quantities\footnote{Note that this is similar to the effective $\theta$
of Gubser and Sondhi, hep-th/0006119.}
\bea
\hat{\mu} &=& \frac{\mu}{\Lambda} \nn \\
\hat{\theta} &=& \theta\Lambda^2\ .
\eea

If we rewrite $A^\dagger A$ in terms of the angular-momentum operators
above, we can then resum the series and find
\be
H = \pi g^2\left[ \log \hat{\theta}\left(\frac{1}{2}N-Q_z\right)
- \log\hat\mu^2 + corrxns. \right]
\ee
The corrections are also polynomial in $A^\dagger A=N-2Q_z$ and so the
Hamiltonian as a whole commutes with both $N$ and $Q_z$. Therefore the
eigenstates are the angular momentum eigenstates, and (since all of the
coefficients in $C$ are positive) the ground state is that anihilated
by $N-2Q_z$. These states have $j=m$, and therefore correspond to all
of the relative angular momentum being in the plane.

The meaning of taking $\Lambda\rightarrow\infty$ in this context is not
fully clear. It seems most natural to leave $\theta$ fixed and so allow
$\hat\theta$ to go to infinity. If one does this, then the two-fermion
effective potential is insensitive to the original value of $\theta$.
A second possibility is to let $\theta\rightarrow 0$ so that $\hat{\theta}$
remains fixed. This parameter is known to be an order parameter in the
striping phase transition of scalar $\phi^4$ theory in noncommutative space,
\cite{GubSond} and it may have relevance in this system as well.

The $\hat{\mu}$ dependence of this Hamiltonian is the ordinary IR divergence
of the Coulomb potential in 2D, and represents an overall constant which is
uninteresting.

\subsection{A check}

Excited states can also be reached by acting with $A^\dagger$ on a ground
state. A good check on the finiteness of this whole matter (as well as on
$\theta$-blindness) is to evaluate $\left<H\right>$ in such a state.
The normalized states are
\be
\frac{1}{\sqrt{2^N N!}} A^{\dagger N}\left|0\right>
\ee
and so
\be
\left<H\right>=g^2\int\frac{d^2 u}{|u|^2} \frac{1}{2^N N!}e^{|u|^2\theta/2}
\left<0\right|A^N e^{-iu A/\sqrt{\theta/2}} e^{-iu^\star A^\dagger/
\sqrt{\theta/2}}A^{\dagger N}\left|0\right>
\ee
where we have reorganized $G$ using the Baker-Hausdorff lemma. This
is then
\bea
\left<H\right>&=&\frac{g^2}{2^N N!}\int \frac{d^2 u}{|u|^2}
\left(i\sqrt{\frac{2}{\theta}}\right)^{2N} e^{|u|^2\theta/2} \left(\partial^N
\bar\partial^N\right) \left<0\right|
e^{-iu A/\sqrt{\theta/2}} e^{-iu^\star A^\dagger/\sqrt{\theta/2}}
\left|0\right> \nn \\
&=& \frac{g^2}{2^N N!}\int \frac{d^2 u}{|u|^2} e^{|u|^2\theta/2}\left(
\partial\bar\partial\right)^N \left[ e^{-|u|^2\theta/2} \left<0\right|G(u)
\left|0\right>\right]\nn \\
&=& \frac{g^2}{2^N N!}\int \frac{d^2 u}{|u|^2} e^{|u|^2\theta/2}\left(
\partial\bar\partial\right)^N e^{-|u|^2\theta}\ .
\eea
The evaluation of the derivatives is a simple matter of combinatorics, and
the integral may be evaluated in polar coordinates as above. The result is
\bea
\left<H\right> &=& 2\pi g^2\sum_{m=0}^N \frac{N!}{(N-m)!(m!)^2}
(-\theta)^m\int du u^{2m-1} e^{-|u|^2\theta/2} \nn \\
&=& 2\pi g^2 \left[ \int \frac{du}{u} e^{-|u|^2\theta/2} + \sum_{m=1}^N
\frac{(-1)^{m-1} N!}{(N-m)!m\cdot m!}\right]\ .
\eea
The first term is infrared divergent but is a constant which may be removed
from the Hamiltonian; it is the analogue of the $\log\hat\mu$ term above.
It is independent of $N$ and so after this removal all of the energies are
finite.

\section{Space momenta}

In the next section, we will derive general rules for the divergence
structure of NCYM with fundamental fermions analogous to the rules for
planar and nonplanar graphs in pure NCYM. While we could do this now,
it is interesting to generalize our considerations to systems where the
fermions have space momenta, and particle-antiparticle creation is
allowed. So this section describes how to add those momenta and get
the general Feynman rules for the theory.

We will begin by solving the first-quantized Dirac equation for fundamental
fermions with kinetic energy. The solutions will be related to SHO
eigenfunctions. The major features which distinguishes these fermions from
those in commutative space are lack of Lorentz-invariance and
the discreteness of the solution set. Once we have these solutions, we
will write down the second-quantized Lagrangian and derive Feynman rules.
With a few simple manipulations, we will show that these rules are almost
identical to those in the commutative theory, except that (1) fermion
momentum integrals are replaced by sums over solutions to the Dirac equation,
and (2) momentum-conserving $\delta$-functions will be replaced with a
smeared function which includes phases. These phases will be key in our
analysis of divergences in the next section.

\subsection{The Dirac equation}

The Dirac equation for fundamental fermions is
\be
(i\not\partial + m)\psi = 0\ .
\ee
Here $\psi$ is an element of the Hilbert space rather than the algebra, so
$\partial$ acts on $\psi$ by multiplication rather than commutation. We pick
metric $\eta=+--$ and $\gamma_0=\sigma_3$, $\gamma_1=i\sigma_1$,
$\gamma_2=i\sigma_2$. We write $\psi(x,t)=e^{-i\omega t}\psi(x)$, and
multiply the equation on the left by $\gamma_0$. Using
$\partial_i = -i\theta^{-1}_{ij}x^j$ this means
\be
\left(\begin{array}{cc}\omega+m & -\sqrt{\frac{2}{\theta}}a^\dagger \\
-\sqrt{\frac{2}{\theta}}a & \omega-m \end{array}\right)\psi = 0
\ee
Setting the determinant equal to zero, we find that this is solved by
\be
\psi = \frac{1}{\sqrt{2\omega(\omega+m)}} \left(\begin{array}{c}
\sqrt{\frac{2n}{\theta}} \ket{n} \\ (\omega+m)\ket{n-1} \end{array}\right)
\ee
where
\be
\omega^2 = m^2 + \frac{2n}{\theta}\ .
\ee
The two signs of $\omega$ correspond, as usual, to particle and antiparticle
solutions, and $\ket{n}$ are SHO eigenstates. We label these states as
$\ket{n\a\pm}$, where $n$ is the SHO level, $\a$ is the spin index, and
$\pm$ is the sign of $\omega$. Off-shell solutions are characterized by
other values of $\omega$.

It is again possible to write these functions as linear combinations of momentum
eigenstates $\ket{z}=G(z)\ket{0}$, which will be useful later but which we
will not do explicitly here; we will simply note that there are functions
such that
\be
\ket{n\a\pm} = \int d^2 z u_{n\a\pm}(z) \ket{z}\ .
\ee

The second quantization of this system proceeds in the usual way. We write
a fermion operator
\be
\Psi = \sum_{n\a} a_{n\a}\ket{n\a+} + b^\dagger_{n\a}\ket{n\a-}
\ee
where $a_{n\a}$ and $b_{n\a}$ are ladder operators satisfying canonical
anticommutation relations. The Hamiltonian density is
\be
{\cal H} = -i\gamma\cdot\nabla + m
\ee
(so that ${\cal H}\ket{n\a\pm}=\pm\gamma^0\omega(n,\a)\ket{n\a\pm}$), and the
overall Hamiltonian is
\be
H = \bar{\Psi}{\cal H}\Psi = \sum_{n\a}\omega(n,\a) \left(a^\dagger_{n\a}
a_{n\a} + b^\dagger_{n\a} b_{n\a}\right)\ .
\ee
(Plus, as usual, an infinite normal ordering constant)

The interactions with the electromagnetic field come from the covariant
derivative, in a term
\be
L_{int} = g \bar\Psi \not A \Psi\ .
\ee
When we quantize $A$, we solve the Klein-Gordon equation for algebra-valued
functions (rather than Hilbert-space valued functions like $\psi$) and as
is well-known the solutions to that are plane-waves $G(z)=e^{-iz\cdot x}$.

\subsection{Feynman rules}

Instead of deriving the Feynman rules for this system in momentum space,
though, we will leave most of the quantities as operators and
take explicit representations only when performing diagrammatic integrals.
(This will make some properties of the system more evident) Thus to a fermion
propagator we associate the factor
\be
\epsffile{eprop.eps} = S_F = \frac{i}{i\not\partial + m}\ .
\ee
(with appropriate placement of $i\epsilon$'s to fix the poles properly)
External fermion lines are associated with terms $\ket{n\a\pm}$ (for incoming
lines) or $\bra{n\a\pm}$ (for outgoing lines.) Internal fermion loops are
associated with bracketing complete expressions in $\bra{n\a\pm}$ and
$\ket{n\a\pm}$ and summing over $(n,\a)$, which is simply a trace over fermion
states; there is also the usual factor of $-1$ for fermion loops coming from
anticommutation. The fermion-fermion-photon vertex comes from expanding $A$
in momentum states and varying the Lagrangian, giving a term
\be
\epsffile{vertex.eps} = -ig(2\pi)\delta(z'_0-z_0-u_0)\gamma^\mu_{\alpha
\dot\beta} G(u)\ ,
\ee
where $\mu$ is the polarization of the photon, $u$ is its momentum, and $\alpha$
and $\dot\beta$ are the spin states of the incoming and outgoing fermion
lines, respectively.

Since vertices and propagators are now operator-valued rather than
number-valued, one should keep in mind the following rules for assembling
them into diagrams. One associates a separate ``channel'' with each fermion
line, which is capped at either end either by bras and kets representing
the external fermion states, or by the internal states which are summed over
to form a trace for internal lines. Then one adds the various relevant
operators $S_F$ and $-ig\gamma G$ in the order in which they appear, starting
from the incoming point and going to the outgoing point. In the case of
internal fermion lines, it does not matter at which point one begins since
the trace is cyclically symmetric.

As an example of how these rules are applied, consider the diagram in
figure~\ref{fig:mess} below. The diagram has three fermionic ``channels'' and
several photon lines. Its value is
\bea
i{\cal M} &=& \int \frac{d^3p d^3 q d^3 r}{(2\pi)^9} (-1) \sum_{l,\g}
\frac{i^3}{q^2 r^2p^2} \bra{n'\a'+}(-ig\gamma^\mu) G(-q) S_F (-ig
\gamma^\nu) G(-p)  \ket{n'\a'+}\nn \\
&&\cdot  \bra{l\g+} S_F (-ig \gamma_\mu) G(q) S_F (-ig \gamma^\rho) G(-r)
\ket{l\g+}\nn \\
&&\cdot \bra{m'\b'-}  (-ig \gamma_\rho) G(r) S_F (-ig \gamma_\nu) G(p)
\ket{m\b-}
\eea

\begin{figure}[hbtp]
\centerline{\epsffile{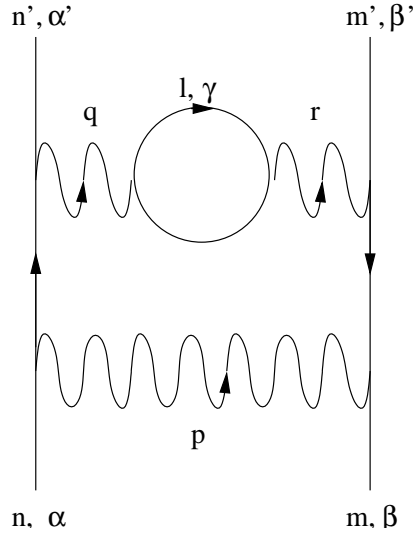}}
\caption{A two-loop diagram}
\label{fig:mess}
\end{figure}

We can reduce these operator rules to ordinary Feynman rules
by using some simple identities. First, from $[\partial_i,G(u)]=-iu_iG(u)$
it follows that
\be
G(u) \frac{i}{i\not\partial + m} = \frac{i}{i\not\partial + \not u + m}
G(u)\ .
\ee

So consider some general Feynman diagram. It will have a photon part which
is possibly very complicated, and for each fermion line it will have a term
that looks like
\be
\bra{n'\a'\pm} S_F G(p_1) S_F \cdots G(p_N) S_F \ket{n\a\pm}\ .
\ee
The outer states will either be fixed states (for external lines) or traced
over (for internal lines). We first push all of the fermion propagators to
the left, giving
\be
\bra{n'\a'\pm} \frac{i}{i\not\partial+m} \frac{i}{i\not\partial+\not p_1+m}
\cdots \frac{i}{i\not\partial + \sum\not p_i + m} G(p_1) \cdots G(p_N)
\ket{n\a\pm}\ .
\ee
Now we insert a complete set of momentum eigenstates after the initial bra,
expressing the state $\ket{n\a\pm}$ in terms of momentum eigenstates. This
gives us
\be
\int dz \bar{u}_{n'\a'\pm}(z) \frac{i}{\not z + m} \cdots \frac{i}{\not z +
\sum\not p_i + m} \bra{z} G(p_1)\cdots G(p_N) \ket{n\a\pm}\ .
\ee
So clearly the propagators are exactly the same as in the commutative case.
The term involving the $G$'s is the analogue of the momentum-conserving
$\delta$-function, which after expanding $\ket{n\a\pm}$ in a momentum basis
is precisely one of the $f_n$'s of equation (\ref{eq:fndef}).

Using these rules, the diagram of figure~\ref{fig:mess} is equal to
\bea
i{\cal M} &=& \int \frac{d^3p d^3 q d^3 r}{(2\pi)^9}(-1)\sum_{l,\g}
\frac{1}{q^2r^2p^2}\left[\bar{u}_{n'\a'+}(z'_1)
(-ig\gamma^\mu) \frac{i}{\not z'_1 - \not q + m}(-ig\gamma^\nu)\right. \nn \\
&&\left.\frac{}{}\cdot u_{n\a+}(z_1) f_2(z_1,z'_1;-p,-q) \right]
\left[\bar{u}_{l\g+}(z'_2) \frac{i}{\not z'_2+m}(-ig\gamma_\mu)\frac{i}
{\not z'_2+\not q + m}\right.\nn \\
&&\left.\frac{}{}\cdot (-ig\gamma^\rho)u_{l\g+}(z_2) f_2(z_2,z'_2;q,-r)\right]
\left[\bar{u}_{m'\b'-}(z'_3)(-ig\gamma_\rho) \frac{i}{\not z'_3+\not r +m}
\right. \nn \\
&&\left.\frac{}{}\cdot (-ig\gamma_\nu) u_{m\b-}(z_3) f_2(z_3, z'_3; p,r)
\right]
\eea
where we have implicitly integrated over the momenta $z_i$. As anticipated,
the only difference between this and the commutative case is the bound-state
nature of the $u_{n\alpha\pm}$ and the replacement of momentum-conserving
$\delta$-functions with $f_n$'s.

\section{Divergence structure}

We wish to find the divergence structure of a general NCYM diagram involving
fermions. This would be the analogue of the rule that planar graphs in
NCYM have only phase factors corresponding to external legs, while nonplanar
graphs have phase factors inside loops and are therefore less divergent.
\cite{BigSus} This followed from introducing a double-line notation in which
lines are associated with their phase factors, and noting that changes in
the diagram which preserve planarity leave all of these factors invariant.
One could then rearrange the diagrams into a tree (out of which external
lines may come), a sequence of tadpoles,
and a nonplanar region. The phase factors from the tree and the tadpoles all
cancel out except for some overall factors on the outer legs, whereas the
phase factors in the nonplanar region cause all loop integrals resulting
therefrom to be strongly damped. The resulting integrals can be evaluated
and contains both the divergences familiar from the commutative theory and
some terms specific to noncommutative Yang-Mills. \cite{NCPert, NCPert2, NCPert3}
We would like to extend this argument to add
single-line diagrams corresponding to fundamentals.
We will see that the rule for fermions is that they never contribute additional
internal phase factors; in fact, at the level of double-line diagrams their
influence can always be replicated by certain planar networks (trees) of
gauge fields.

The argument is as follows. We first consider the issue of momentum
nonconservation. If in any subdiagram of a given diagram the wobble
($q+z-z'$ in (\ref{eq:fndef})) is nonzero, that diagram contains both
Gaussian damping factors and
phase factors containing both $q$ and fermion momenta. Therefore, the leading
divergences come from terms where there is no nonconservation, and so the
only effect of the $f_n$ is to introduce phase factors at each fermion-photon
vertex.

Because this is the only remaining difference from the commutative
case,\footnote{When compared to a commutative system where the fermions are
in appropriate bound-state wavefunctions} from here on out we will be
concerned only with phase factors and can therefore draw simplified diagrams
where photons are represented by
double lines, fermions by single lines, and each vertex is associated
simply with its phase factor $e^{iz'\wedge z}$. (cf. (\ref{eq:f1val}) when
$q=0$) We draw shaded circles for
each vertex to make their location more clear in the figures.

The procedure for evaluating phase contributions to divergences is as
follows. Consider an arbitrary diagram with photon and fermion lines.

\begin{enumerate}
\item{First, ``purely internal'' networks of fermions can be integrated
out. These are defined as networks of fermion lines which do not include
any external fermions. Since fermions can only attach at $\psi\psi\gamma$
vertices, such networks must form polygons with $N$ external photon legs.
If the momenta are as shown in figure (\ref{fig:polygon}) below, the
diagram will contain a term
\bea
\lefteqn{\int dz dz' \bra{n'\a'+}\noket{z'}\bra{z}\noket{n\a+}
\exp i\left(\sum_{j<i} q_i\wedge q_j + (z+z')\wedge \left(\sum q_i\right)
- z\wedge z'\right)\cdots} \nn \\
&=&  \int dz dz' \bra{n'\a'+}\noket{z'}\bra{z}\noket{n\a+}
\exp i\left(\sum_{j<i} q_i\wedge q_j - z\wedge z'\right)\cdots \ .
\eea
since the wobble $\sum q_i+z-z'=0$. This diagram is clearly most divergent
when $z\wedge z'=0$, where it becomes equal to its commutative-space value
times an overall phase factor
\be
\exp i\sum_{j<i} q_i\wedge q_j\ .
\ee
Terms with phase factors contribute only finite amounts, since the exponential
phase factor can cancel any power-law divergence.

\begin{figure}[hbtp]
\centerline{\epsffile{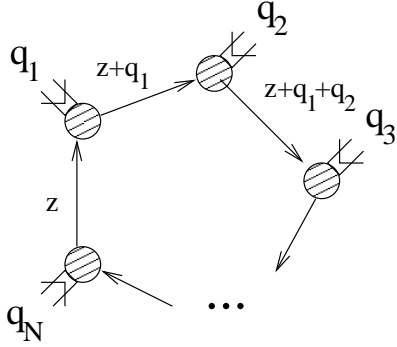}}
\caption{An internal fermion network.}
\label{fig:polygon}
\end{figure}

This phase factor has two important properties. First, it does not involve
the loop momentum at all, so there is no damping of this loop integral
relative to the commutative theory. Thus this loop is as divergent as usual.
Second, the phase factor for this effective $N$-photon vertex is exactly
equal to the phase factor one would get from a planar subdiagram (a tree) of pure
NCYM with $N$ external legs. We can therefore integrate out this fermion
loop, with the diagram receiving whatever divergences would be appropriate
in the commutative theory, and replace it with such an effective tree
for the purpose of phase analysis.
}
\item{Now we consider external fermion legs. An external fermion can attach
to the rest of the diagram in one of two ways: (fig.~\ref{fig:extferm})

\begin{figure}[hbtp]
\epsfxsize=\textwidth
\centerline{\epsffile{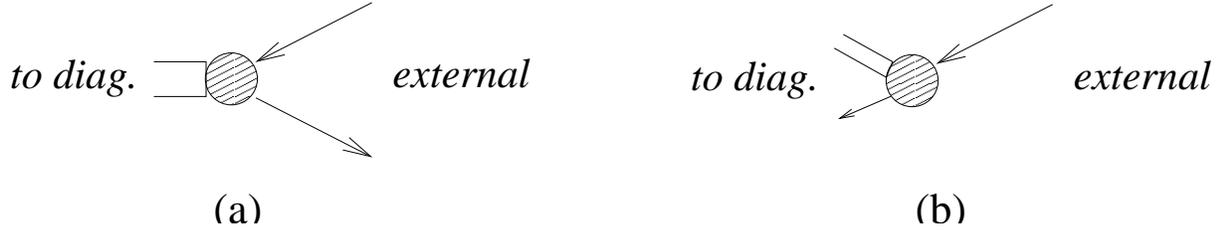}}
\caption{External fermion attachments.}
\label{fig:extferm}
\end{figure}

If it attaches by (b), then there is a single fermion line entering the
rest of the diagram which again must attach via either (a) or (b). This
process repeats indefinitely until it is cut off by a connection of type
(a). Therefore all external fermion connections must take the form of
fig.~\ref{fig:extfermvert}.

\begin{figure}[hbtp]
\epsfysize=2.in
\centerline{\epsffile{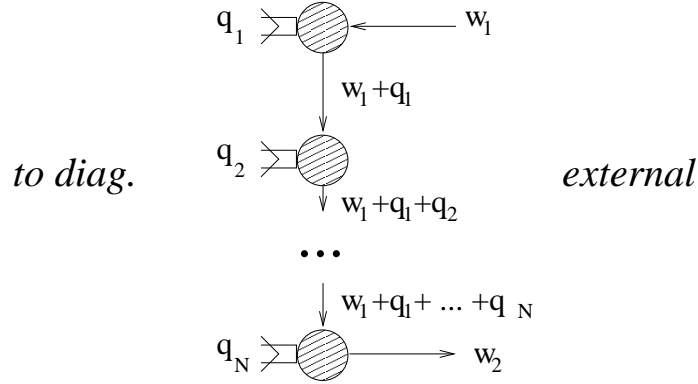}}
\caption{All external fermion attachments must ultimately have this form.}
\label{fig:extfermvert}
\end{figure}

This is an effective $N$-photon 2-fermion vertex, whose phase may be
easily calculated. Using the momentum conventions given in the figure, the
phase is

\bea
\phi &=& \exp i\left[(w_1+q_1)\wedge w_1 + (w_1+q_1+q_2)\wedge(w_1+q_1) +
\cdots\right.\nn \\
&& \left.+ (w_1+q_1+\cdots q_N)\wedge(w_1+q_1+\cdots q_{N-1})\right] \nn \\
&=& \exp i\left[q_1\wedge w_1 + q_2\wedge(w_1+q_1) + \cdots q_N\wedge
(w_1+q_1+\cdots q_{N-1})\right] \nn \\
&=& \exp i\left[w_2 \wedge w_1 + \sum_{j<i\le N} q_i\wedge q_j\right]
\eea
where we have used the fact that $w_2 = w_1 + \sum q_i$. But this phase
is precisely the phase that would come from an $(N+1)$-photon vertex
attached to a single fermion-fermion-photon vertex, so we can replace this
effective vertex with precisely that. (As before, the $(N+1)$-photon vertex
can then be reduced to a planar network of NCYM vertices)
}
\item{We have therefore eliminated all internal fermion lines in favor of
planar networks of NCYM vertices, and attached all external fermion lines
with fermion-fermion-photon vertices where both fermions are external. It
is therefore possible to now perform the ordinary channel-swapping moves
of NCYM to rearrange the purely photonic part of the diagram into a tree,
some tadpoles, and a nonplanar part, with some of the external photon
lines now ``capped'' with a fermion-fermion-photon vertex. From the
ordinary evaluation of divergences, we know that the nonplanar part will
be suppressed by phase factors inside loops, and the divergences will be
dominated by diagrams whose photonic part is purely planar. Since internal
fermion lines always contribute planar NCYM effective diagrams, we find
that fermions never decrease the degree of divergence in a graph.}
\end{enumerate}

\medskip

The theory of fundamentals on the noncommutative plane is therefore a
straightforward extension of ordinary NCYM theory. The behavior is not
qualitatively surprising, reducing simply to the combination of bound-state
wavefunctions and the momentum nonconservation which we have seen in several
forms. The major open question at this point is the extent to which this
can be used to construct physically interesting and realistic models of
systems in strong background fields, which will hopefully further elucidate
the meaning of spacetime noncommutativity in physics.

\bigskip

\centerline{\sc\bf Acknowledgements}

The author wishes to thank Leonard Susskind for extensive discussions and
assistance on this work, especially on the subjects of sections 2 and 4.
Thanks also to Nelia Mann for reading a preliminary version of this file and
making numerous helpful suggestions. This work was supported by the National
Science Foundation (NSF) under grant PHY-9870015.

\end{document}